\documentclass[sigconf]{acmart}

\renewcommand\footnotetextcopyrightpermission[1]{} 
\usepackage{amsmath}
\usepackage{tabularx}

\usepackage{graphicx}
\usepackage{subcaption}

\usepackage{balance}

\begin{document}
\title{Optimal Sabotage Attack on Composite Material Parts} 

\author{Bikash Ranabhat, Joseph Clements, Jacob Gatlin, Kuang-Ting Hsiao, Mark Yampolskiy}
\affiliation{%
  \institution{University of South Alabama}
}

\begin{abstract}

\emph{Industry 4.0} envisions a fully automated manufacturing environment, in which computerized manufacturing equipment---Cyber-Physical Systems (CPS)---performs all tasks.
These machines are open to a variety of cyber and cyber-physical attacks, including sabotage. 
In the manufacturing context, sabotage attacks aim to damage equipment or degrade a manufactured part's mechanical properties.  
In this paper, we focus on the latter, specifically for composite materials.
Composite material parts are predominantly used in safety-critical systems, \emph{e.g.,} as load-bearing parts of aircraft. 
Further, we distinguish between the methods to compromise various manufacturing equipment, and the malicious manipulations that will sabotage a part.
As the research literature has numerous examples of the former, in this paper we assume that the equipment is already compromised; our discussion is solely on manipulations.

We develop a simulation approach to designing sabotage attacks against composite material parts. 
The attack can be optimized by two criteria, minimizing the ``footprint'' of manipulations. 
We simulate two optimal attacks against the design of a spar, a load bearing component of an airplane wing.
Our simulation identifies the minimal manipulations needed to degrade its strength to three desired levels, as well as the resulting failure characteristics. 
Last but not least, we outline an approach to identifying sabotaged parts.

\end{abstract}


\maketitle

\section{Introduction}

\emph{Industry 4.0} and \emph{Factory of the Future} movements envision fully automated manufacturing environments, in which computerized manufactured equipment-- Cyber-Physical Systems (CPS)-- perform all tasks.
These are vulnerable to a  variety of cyber and cyber-physical attacks, as has been shown for CPS. 
However, as opposed to equipment sabotage attacks like Stuxnet~\cite{falliere2011w32}, manufacturing sabotage attacks might aim to degrade the mechanical properties of the manufactured part.
This can lead to the destruction of a system employing such a part.
The \emph{dr0wned} study~\cite{belikovetsky2017dr0wned} provides an experimental proof of these attacks; in the study, researchers sabotaged a replacement propeller for a quad-copter UAV, leading to the quad-copter's destruction after the propeller broke mid-flight.

\begin{figure}[tbp]
	\centering
		\includegraphics[width=0.45\textwidth]{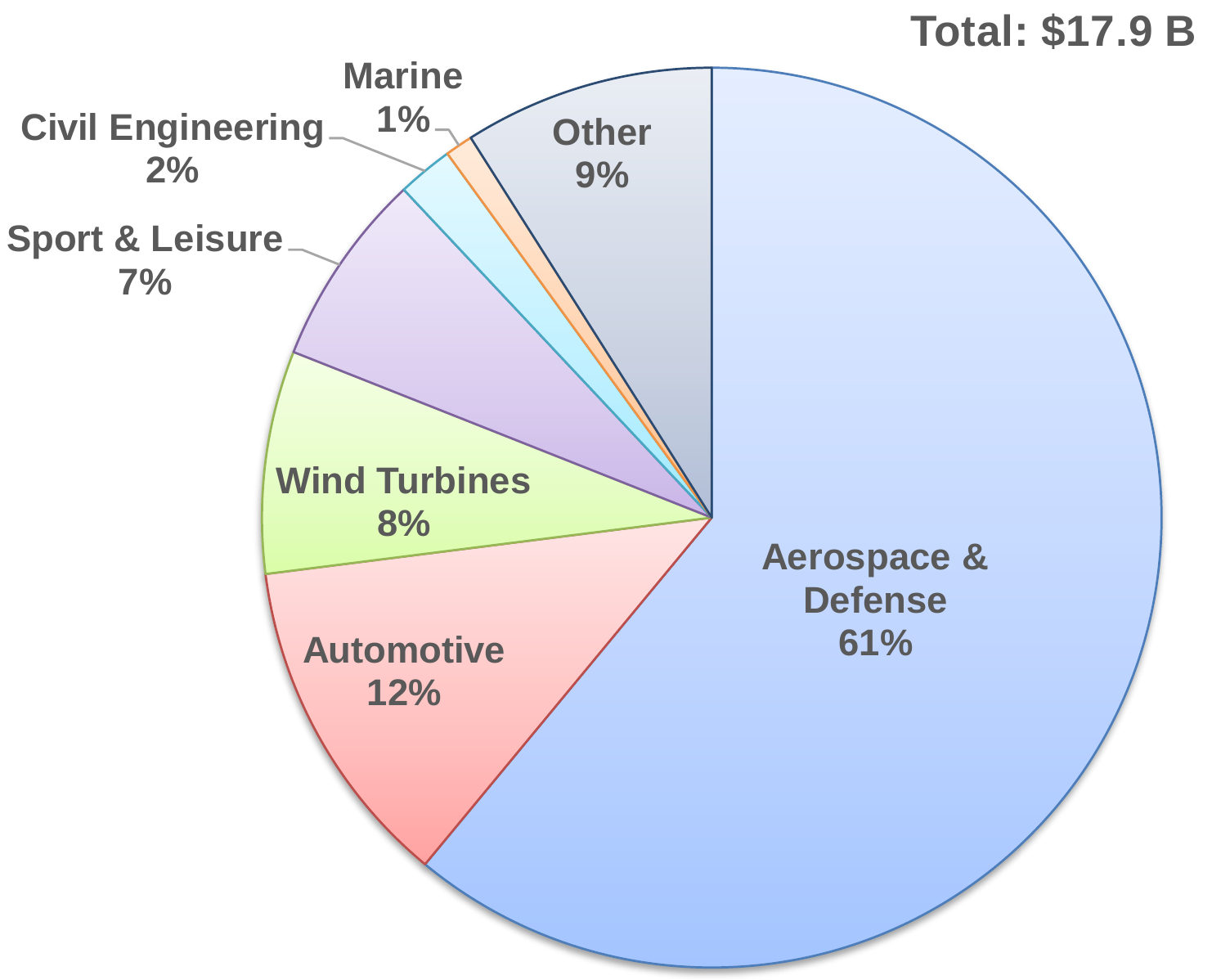}
	\caption{Composite Material Market, 2016 (Source:~\cite{avk2016composites})}
	\label{fig:CFRP-report-2016}
\end{figure}

In this paper we focus on composite materials, which are commonly used in safety-critical systems. 
The aerospace industry was an early adopter of this innovation, and composite use in aerospace has increased drastically over time.
For instance, the Boeing 787 is 50\% composite by weight~\cite{boeing787url, kim2014proceedings}. 
The defense industry also favors composite materials.
Composite application areas include aerospace structures (tails, wings and fuselages), missile components (rocket motor cases, nozzles, igniters, inter-stage structures), boat construction, bicycle frames, and racing car bodies (see Figure~\ref{fig:CFRP-report-2016}). 

Most large composite parts are manufactured through automated processes, instead of the traditional labor-intensive hand-layup processes that are still popular for small parts.
Incorporating automated processes like automatic fiber placement, automated tape layup, filament winding, etc. for manufacturing composites raise security concerns. 
Increased computerization and a key place in the aerospace and defense industries make composite material manufacturing an attractive target for cyber- and cyber-physical attacks. 
While we are not aware of any prior publications discussing sabotage attacks on composite material parts, the possibility of sabotage attacks has already been empirically proven for \emph{Additive Manufacturing} (AM, a.k.a. \emph{3D Printing}) with plastics and metals. 
In AM, it has been shown that sabotage attacks can degrade a part's mechanical strength~\cite{sturm2014cyber, yampolskiy2015security, zeltmann2016manufacturing} or fatigue life~\cite{belikovetsky2017dr0wned}.

The contributions of this paper are as follows.
We demonstrate that sabotage attacks on composite material parts are possible, define two categories of optimal sabotage attacks on composite material parts, and present a simulation tool developed to calculate both categories of optimal attacks. 
We further discuss the properties of these optimal attacks, and outline approaches that might be used to identify sabotaged composite material parts. 

This paper is structured as follows. 
In Section~\ref{sec:background}, based on the current state of the art, we present the composite material ma\-nu\-fac\-tu\-ring foundation of our approach. 
After discussing the considered threat model and related assumptions in Section~\ref{sec:threat}, 
in Section~\ref{sec:optimal-attacks} we define both categories of optimal sabotage attacks and outline the algorithms used to calculate them, for a given part and degradation requirements. 
We present and discuss the properties of the achieved optimal attacks in Section~\ref{sec:simulation-results}. 
We outline possible approaches to identify sabotage attacks in Section~\ref{sec:detectability}. 
After discussing sabotage attacks identified for AM in Section~\ref{sec:SOTA}, 
we conclude this paper with a brief overview of the presented work and conclusions from it.

\section{Background}
\label{sec:background}

In this section, we provide background material for composite materials and their manufacturing relevant to this paper. 
In Appendix
~\ref{sec:appendix} we summarize a mathematical method that is well-established in materials science for the calculation of a composite material's mechanical properties.

\subsection{Carbon Fiber Reinforced Polymer (CFRP)}

\begin{figure}[tbp] 
	\centering
		\includegraphics[width=0.5\textwidth]{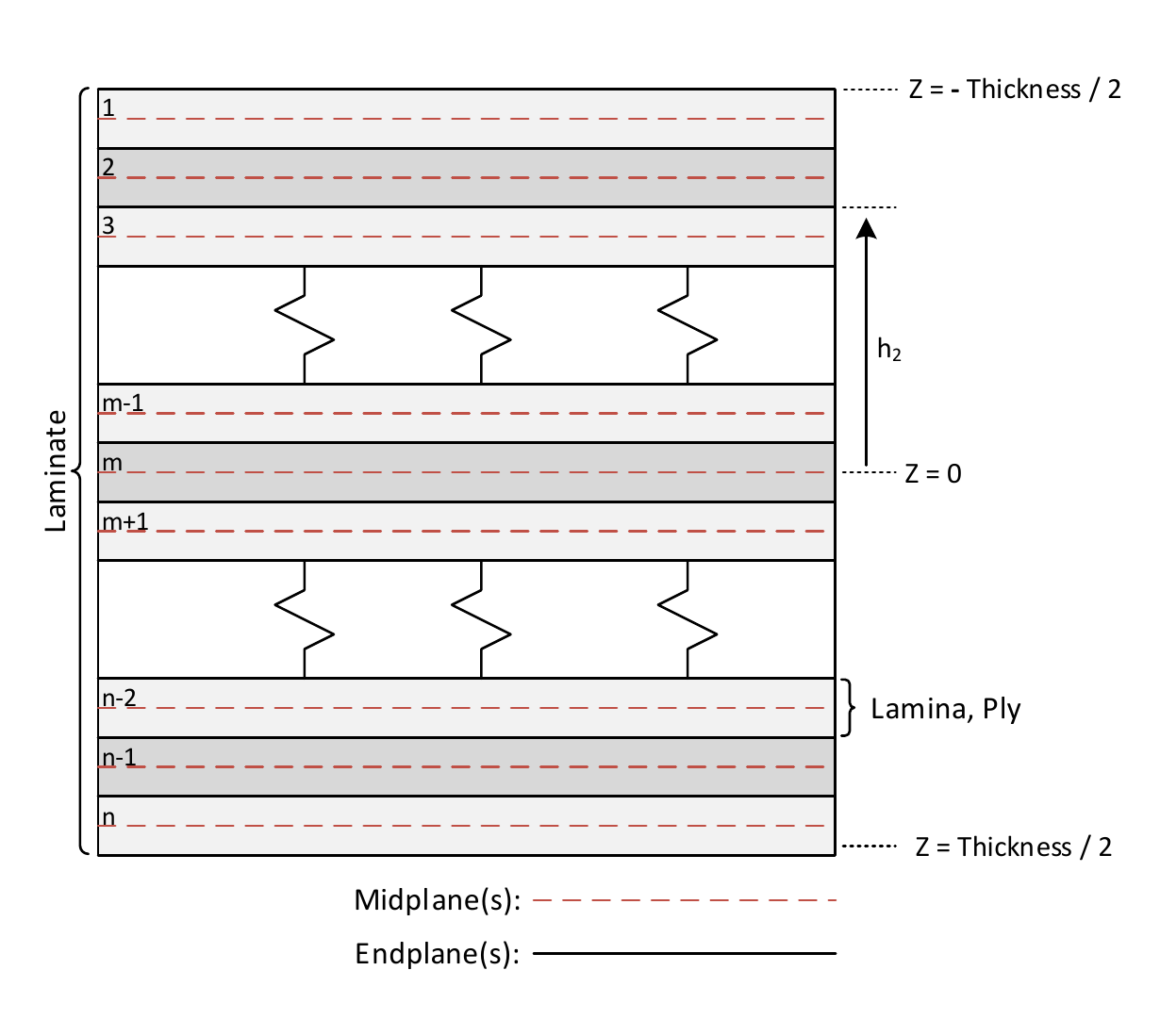}
	\caption{Laminate, Lamina, Mid- and End-Planes}
	\label{fig:LaminatePlyIllistration}
\end{figure}

\emph{Composite materials} are the combination of two or more materials with different sets of properties. The composite benefits from the best properties of each of the combined materials or achieves a new set of properties which are unique and cannot be achieved in either of the individual materials. Despite their high cost, composite materials have gained popularity in high performance products that need to be lightweight but able to bear large loads. 

This paper primarily considers \emph{Carbon Fiber Reinforced Polymer} (CFRP), a composite that is very popular in the aerospace and automotive industries because of its high strength-to-weight ratio. 
CFRP derives its strength from sheets of carbon fiber that are bound together with a polymer-like epoxy. 
Individual sheets of carbon fiber are referred to as \emph{lamina} or \emph{plies}, and the complete composite as a \emph{laminate} (see Figure~\ref{fig:LaminatePlyIllistration}).

\begin{figure}[tbp]
	\centering
		\includegraphics[width=0.5\textwidth]{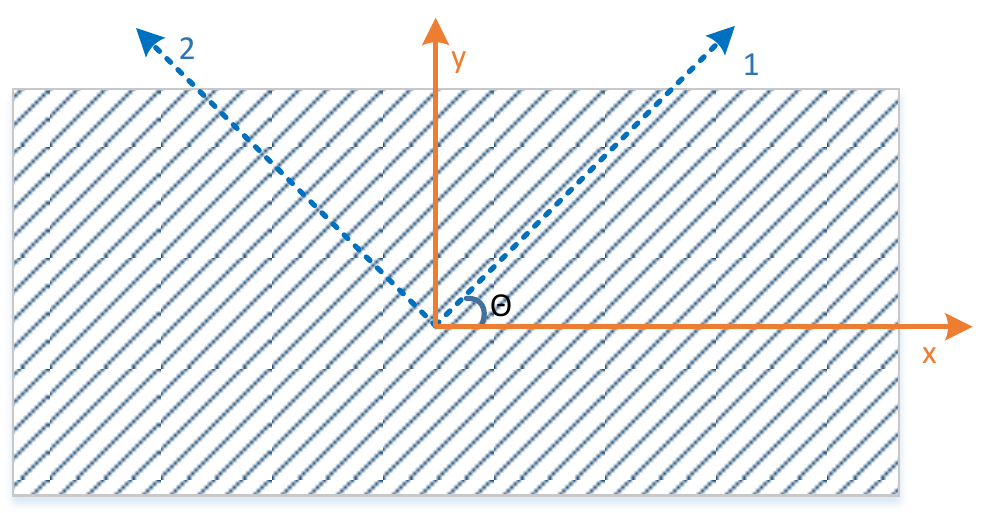}
	\caption{1/2 and X/Y Axes in Composite Materials}
	\label{fig:12XY-Axes}
\end{figure}

Each individual carbon fiber lamina shows strong anisotropic properties, i.e., mechanical characteristics like strength and stiffness are different in different directions. 
To control this anisotropy, laminas are placed in the laminate at different angles. 
Figure~\ref{fig:12XY-Axes} shows the local and global axes of an angle lamina. Vectors 1 and 2 represent the local coordinate axes whereas x and y are the global coordinate axes. 
Vector 1 lies along the fiber length direction, which represents longitudinal direction. Vector 2 is normal to the fiber length and represents transverse direction for the lamina. 
The global coordinate system (x-y) is related to the local coordinate system by the angle ($\Theta$).

In Appendix~\ref{sec:appendix} we have summarized the well-established equations commonly used to calculate the mechanical properties of CFRP.

\subsection{Computer Aided Manufacturing of Composite Material}

The composites industry is moving from hand layup to automated processes. \emph{Automated Tape Layup} (ATL) and \emph{Automated Fiber Placement} (AFP) are the current automated additive manufacturing techniques for composites in the industry. ATL and AFP are performed by Computer Numerical Control (CNC) machines with programmable axis movement, axis drives, and a printing head. They place laminated tapes and tow layer by layer to produce a specific part. 

\subsubsection{Automated Tape Layup}

ATL can be considered an Additive Manufacturing process used to produce large simple composite structures. ATL can lay prepared tape and continuous fabric strips ranging from 75 to 300 mm wide onto a flat surface in different orientations~\cite{sloan2008atl, dirk2012engineering, beakou2011modelling}. ATL heads are comprised of spools of tape, a winder, winder guides, a compaction roller, a tape cutter, and a position sensor. ATL composite production starts by depositing a prepared tape onto a mold using a silicone roller. Subsequent layers are deposited according to CNC paths defining the part geometry. Once the layup is completed, the tape is cut automatically by rotating or pinching blades. The speed of machine layup depends on the complexity of the part. 
There is a reported 70-85\% reduction in man hours in the aerospace industry due to the implementation of ATL techniques~\cite{grimshaw2001advanced}.

\subsubsection{Automated Fiber Placement}

The AFP process places bands of material consisting of 2-32 individual tows or slit tape using low tension and compaction pressure~\cite{debout2011tool}. 
The tows are usually 1/8'' to 1/4'' wide. These tows can be laid at their own speed and start/stop along the length of the band at the desired locations. The use of robots allows the AFP process to be highly controllable and repeatable.
For fabrication, creel systems are loaded with tows of composite tape. During band placement, individual tows are fed and cut as programmed by the program controller. The tows input to the system pass in front of a heat source (laser, hot gas torched) and under the consolidation device. 
The heat melts the tape, causing it to stick to the substrate 
when pressed down by the consolidation device. Once a band is placed, the tows are cut and the robot head returns to place the next band. The process repeats band-by-band, until a ply is complete, and then ply-by-ply until the final part is produced.

\subsection{Quality Assurance}

Optical systems are being integrated by most manufacturers for inline quality control during the AFP process~\cite{dirk2012engineering}. 
Fiber orientation 
is an important parameter to monitor during manufacturing; 
sudden changes in fiber orientation indicate defects such as gaps, overlaps, twisted tows and contamination with foil. The strength and stiffness of the composite laminate greatly depends on the orientation sequence of the plies. Axial load is taken by ${0^{\circ}}$ plies, shear load by ${\pm{45}^{\circ}}$ plies and transverse load by ${90^{\circ}}$ plies. Fiber orientation is therefore a major criterion for quality assurance (QA).

The optical system, \emph{Automated Ply Inspection} (API), was developed by \emph{Flightware} for NASA's two AFP machines. It consists of a sensor mounted in the AFP head~\cite{maass2015progress}. A laser line is projected on the layup surface. A commercial laser line scanner acquires the optical measurement of the layup surface in a two-dimensional profile, usually comprised of 1000-1500 discrete points. Automated scanning generates a very accurate and dense point cloud that includes the minute details of the layup surface. The position of the sensor in space at any instant is known so 2D data points from the sensor could be converted into 3D coordinates in the part coordinate system. Thus, lines of data points can be obtained hundreds of times per second. Layup features of interest such as overlaps, edges, gaps, splices, orientation etc. are embedded in the point cloud data and layup surface points are plotted in 3D space. Comparing the actual ply edge location to an expected ply edge location defined on the part coordinate system helps determine the laydown accuracy of the machine. The whole process is automated by the Flightware software called \emph{Feature Detection}.


\section{Threat Model \& focus of the Paper}
\label{sec:threat}

While we are not aware of any prior research on composite materials security, 
composite manufacturing is quite similar to Additive Manufacturing (AM), a.k.a. 3D Printing.
For attacks on or with AM, one should distinguish between \emph{attack vectors}, \emph{compromised elements}, \emph{manipulations}, and \emph{effects}~\cite{yampolskiy2016using}. 
A variety of \emph{attack vectors} can be used to compromise one or more elements of the manufacturing process. 
The \emph{compromised element(s)}, their roles in the process, and the degree to which an adversary can control these element(s) determine which \emph{manipulations} an adversary can perform. 
These manipulations, in conjunction with the type of manufacturing process, manufacturing equipment, source materials, and object application area, determine the achievable \emph{effects}. 
Only a fraction of the achievable effects intersect with the adversary's goals. 
The intersection of attack effects and adversarial goals are \emph{attack targets} (or \emph{threats}).
We consider these points valid for composite materials manufacturing.

For AM, two major security threats identified in the research literature are intellectual property theft~\cite{yampolskiy2014intellectual, faruque2016acoustic, faruque2016forensics, song2016my, hojjati2016leave} and sabotage attacks~\cite{sturm2014cyber, yampolskiy2015security, zeltmann2016manufacturing, belikovetsky2017dr0wned}. 
These threats are also present in composite materials. In this paper, we focus explicitly on the latter threat category. 
Furthermore, in the manufacturing context, sabotage attacks aim to 
degrade a manufactured object's material properties, damage manufacturing equipment, or damage/contaminate the environment of either the manufactured part or the manufacturing equipment~\cite{yampolskiy2016using}. 
This paper focuses solely on sabotage attacks aiming to degrade a manufactured part's mechanical properties. 

Numerous real-world as well as academic attacks have proven that a broad variety of classical cyber attack vectors can be used to compromise computerized elements of a manufacturing process. 
The most prominent real-world example remains the Stuxnet attack~\cite{falliere2011w32}, which has compromised both SCADA computers and manufacturing equipment firmware (a uranium enrichment centrifuge). 

Examples from the closely related AM security literature 
(we provide a more detailed description in Section~\ref{sec:SOTA}) 
are also numerous. 
Considering attack vectors, researchers have 
complained about the lack of the security features associated with design files or design file transfer~\cite{turner2015bad}, 
identified that the employed software and firmware have numerous vulnerabilities~\cite{moore2016vulnerability}, exploited weaknesses in the communication protocol to highjack the session~\cite{do2016data}, or used social engineering to convince a user to open a malicious e-mail attachment~\cite{belikovetsky2017dr0wned}.
Compromised elements include computers that store design files and control the manufacturing process~\cite{sturm2014cyber, belikovetsky2017dr0wned}, communication networks that connect these computers with the robotized manufacturing equipment~\cite{do2016data}, the firmware installed on the equipment~\cite{xiao2013security, moore2017implications}, or even an in-situ quality control system~\cite{slaughter2017how}.
These compromised elements can stage a variety of manipulations, including tampering of the design files~\cite{sturm2014cyber, belikovetsky2017dr0wned}, manufacturing parameters~\cite{yampolskiy2015security, zeltmann2016manufacturing, moore2017implications}, manufacturing process status information~\cite{slaughter2017how}, and altering communication timings or power supply~\cite{pope2016hazard, yampolskiy2017evaluation}. 

The fundamental difference between AM and composite materials security lies in 
the parameters that determine the mechanical properties of the manufactured part.
This also determines which malicious manipulations are available to an adversary to sabotage a part's quality.

Based on the prior discussion, we consider the following threat model:

\begin{itemize}
	\item Supported by prior work, we assume that one of the computerized elements in the manufacturing process is compromised. It can be a PC that stores design files and controls the overall manufacturing process, a computer network that connects this PC and the roboticized 
  ATL or AFP machine, 
	or the firmware on this machine. 
	
	\item In this paper, we focus on distinct manipulations that are characteristic to all composite material parts--the orientation of individual plies in a laminate.
	This can be performed, e.g., by modification of the design files on a compromised computer. 
	Other malicious manipulations like changing the temperature-profile during the curing process are possible, but are out of scope for this paper. 
	
	\item Functional parts are typically designed with extreme tolerable operational conditions in mind. We assume that an adversary knows this and can discern the operational conditions based on the design alone. Furthermore, we assume that the adversarial goals can be tied to the (discernible) operational conditions. 
	
	\item Last but not least, the example of the--supposedly state-actor staged--Stuxnet sabotage attack also shows that adversaries will go to great lengths for attack stealthiness while still achieving their goals. We therefore assume that an adversary will be willing to minimize the malicious manipulations (in the case considered in this paper, changes of the individual ply orientations) while still achieving their stated goal (in the considered case, reducing the mechanical properties to the level below discernible operational conditions).
\end{itemize}

\section{Optimal Sabotage Attacks}
\label{sec:optimal-attacks}

The mechanical properties of a composite material directly depend on several factors. The most important are the number of plies, 
the material of individual plies, the orientations of individual plies, the presence of defects like gaps or overlaps, and bonding between plies. This paper alters only ply orientation, holding all other parameters constant. 

For part degradation sabotage attacks achieved through ply orientation, we define two categories of 
optimization criteria:
(i) minimal amount of orientation deviation ($\Delta\Theta$) for all 
plies, and (ii) minimal number of plies whose orientation is changed.
We consider attacks that satisfy these criteria as \emph{optimal}, because they minimize the ``footprint'' of the attack while achieving the specified degradation characteristics.

Based on the mathematical foundation described in Appendix~\ref{sec:background:math}, we developed a MATLAB program that takes as input an unaltered 
composite material design (i.e., a sequence of plies, including their material and orientation) and calculates both optimal attacks. The simulation program consists of three functions, described in more detail in this section. 

The first calculates a strength ratio for each ply in a laminate; it also determines the sequence in which the layers will fail as well as the corresponding forces. The remaining two functions use the first as a sub-routine and calculate two 
categories of optimal sabotage attacks using defined optimization strategies.


\subsection{Helper Function: Calculate Max Force and Ply Failure Sequence}
\label{sec:optimal-attacks:helper-function}

Please note that all equations referenced in this section are described in Appendix~\ref{sec:appendix}; these equations are well established in materials science. 

\begin{figure}[tbp] 
	\centering
		\includegraphics[width=0.5\textwidth]{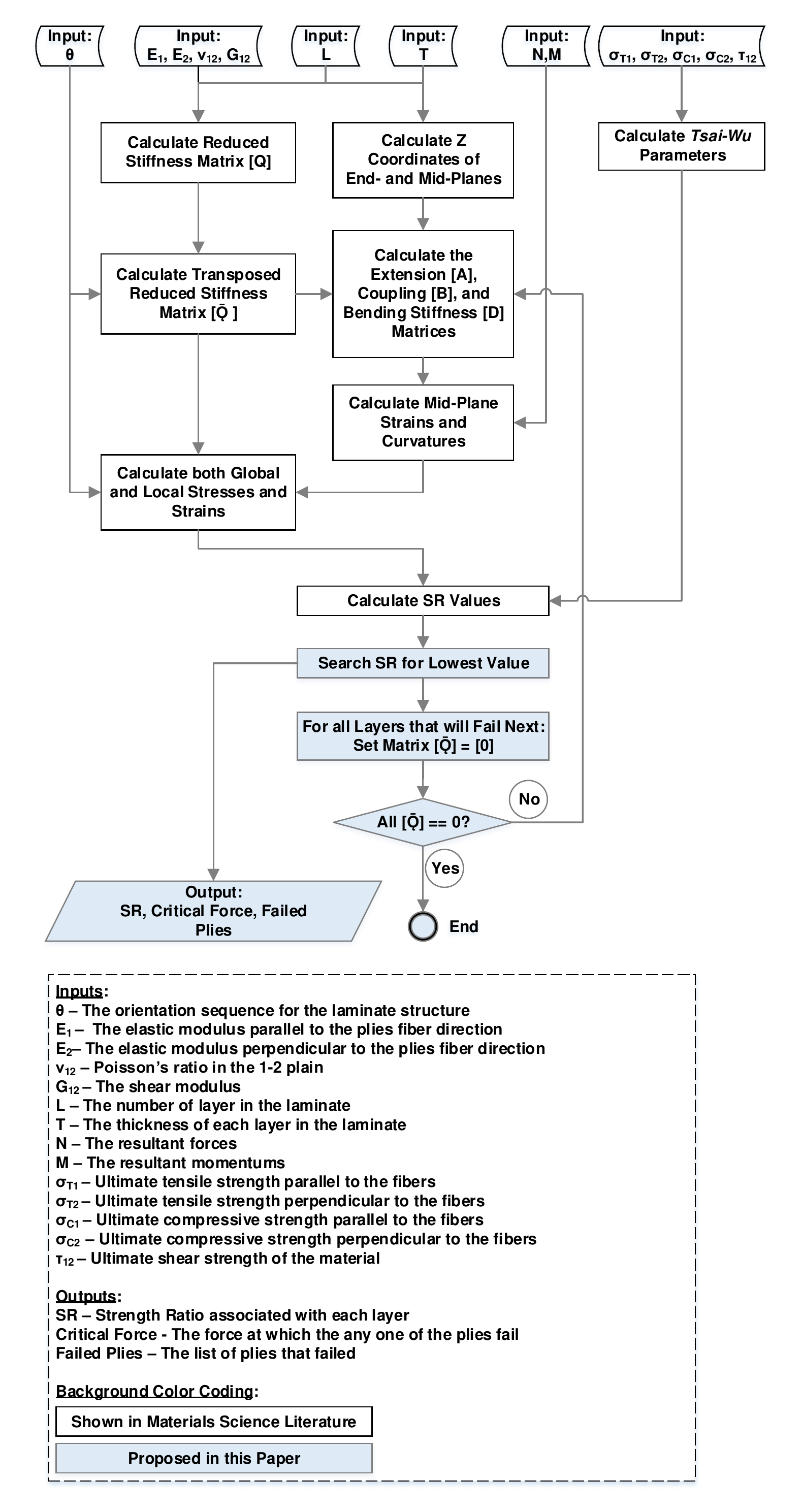}
	\caption{Calculate Max Force and Ply Failure Sequence}
	\label{fig:OptimalAttackCalculation_HelperFunction}
\end{figure}

The flowchart in Figure~\ref{fig:OptimalAttackCalculation_HelperFunction} summarizes an algorithm that is used as a core component in the calculation of both attack types described in this section. This function is based on the algorithm for the \emph{Strength Ratio} ($SR$) calculation as described in Ramsaroop et al.~\cite{ramsaroop2010using}; our modifications are indicated with a darker (sea blue) background.

The algorithm takes as its inputs the material properties, the details of the laminates, a vector of resultant forces, and moments indicating how the forces are distributed. The output of the algorithm is an array of $SR$ for all plies and the sequence of forces at which plies fail.

We start with the input values of the Young's modulus, Poisson's ratio, shear  modulus (${E_1}$, ${E_2}$, ${\nu_{12}}$, ${G_{12}}$), and number of layers. We use these values to calculate elements of the reduced stiffness matrix ${[Q]}$ according to the Equation~\ref{eq:3}. This matrix and the array of the orientation (${\Theta}$) are then used to calculate the array of transposed reduced stiffness matrices ${[\overline{Q}]}$ 
for each ply; the calculation is performed according to the Equations~\ref{eq:5} and~\ref{eq:4}. 

We use the information about layer thickness and the number of layers to calculate the $Z$ coordinates of mid- and end-planes for each ply (see Figure~\ref{fig:LaminatePlyIllistration}). 
The sign convention used here is positive for the downward layer and negative for the upward layer, with the middle layer as reference.

We then use the previously computed transposed reduced stiffness matrix ${[\overline{Q}]}$ 
and the coordinates of end-planes to calculate the extensional stiffness matrix ${[A]}$, coupling stiffness matrix ${[B]}$, and bending stiffness matrix ${[D]}$ according to Equations~\ref{eq:9},~\ref{eq:10}, and~\ref{eq:11}, respectively. 
These matrices, the resultant forces $N$, and resultant moments $M$ are then used to calculate mid-plane strains  ${\epsilon^0}$ 
and curvatures $k$ according to Equation~\ref{eq:7}. 

All prior calculated values are used to determine the global and local stresses and strains. 
For the calculation of global strains Equation~\ref{eq:8} is used. We use Equation~\ref{eq:2} to calculate global stresses. To calculate local strain and stress, Equation~\ref{eq:6} is used, which relies on the values of the global strain and stress. 

The components of Tsai-Wu failure theory are calculated using Equations~\ref{eq:13-1} through~\ref{eq:13-3}; for this, the ultimate tensile stresses ${(\sigma^T_{1,2})_{ult}}$, the ultimate compressive stresses ${(\sigma^C_{1,2})_{ult}}$, and the ultimate shear stress ${(\tau_{12})_{ult}}$ are used. The Tsai-Wu parameters together with the previously determined local stress are then used to calculate the laminate's ${SR}$ values, one per ply. 

Up till this point, the described calculations are derived from the approach layout in~\cite{ramsaroop2010using}. 
The remaining calculations have been introduced in order to accommodate the specific needs of our algorithms which determine optimal attacks. 

After the ${SR}$ values are calculated, the array is searched to find the plies with the lowest ${SR}$ value. These are the plies which fail first. 
These plies fail at the force equal to the product of their ${SR}$ value and the resultant forces $N$ and moments $M$.
We remove these plies from the further considerations by setting their transposed reduced stiffness matrix ${[\overline{Q}]}$ 
equal to ${[0]}$. 

If not all transposed reduced stiffness matrices ${[\overline{Q}]}$ 
have been set to ${[0]}$, this indicates that at least one ply has not failed yet and can withstand the applied force.  
In this case, we re-enter the algorithm at the stage when the extensional stiffness ${[A]}$, coupling stiffness ${[B]}$, and bending stiffness matrices ${[D]}$ are calculated. Otherwise, all plies have failed and we can exit the algorithm. 

During the algorithm, the following output values are calculated and updated when necessary. The ``Critical Force'' is implemented as an array of forces under which one or more plies will fail. With every element of the array, a list of plies is associated that will fail under this force. 


\subsection{Optimal Attack, Type One: Minimal Changes per Individual Ply}

\begin{figure}[tbp] 
	\centering
		\includegraphics[width=0.5\textwidth]{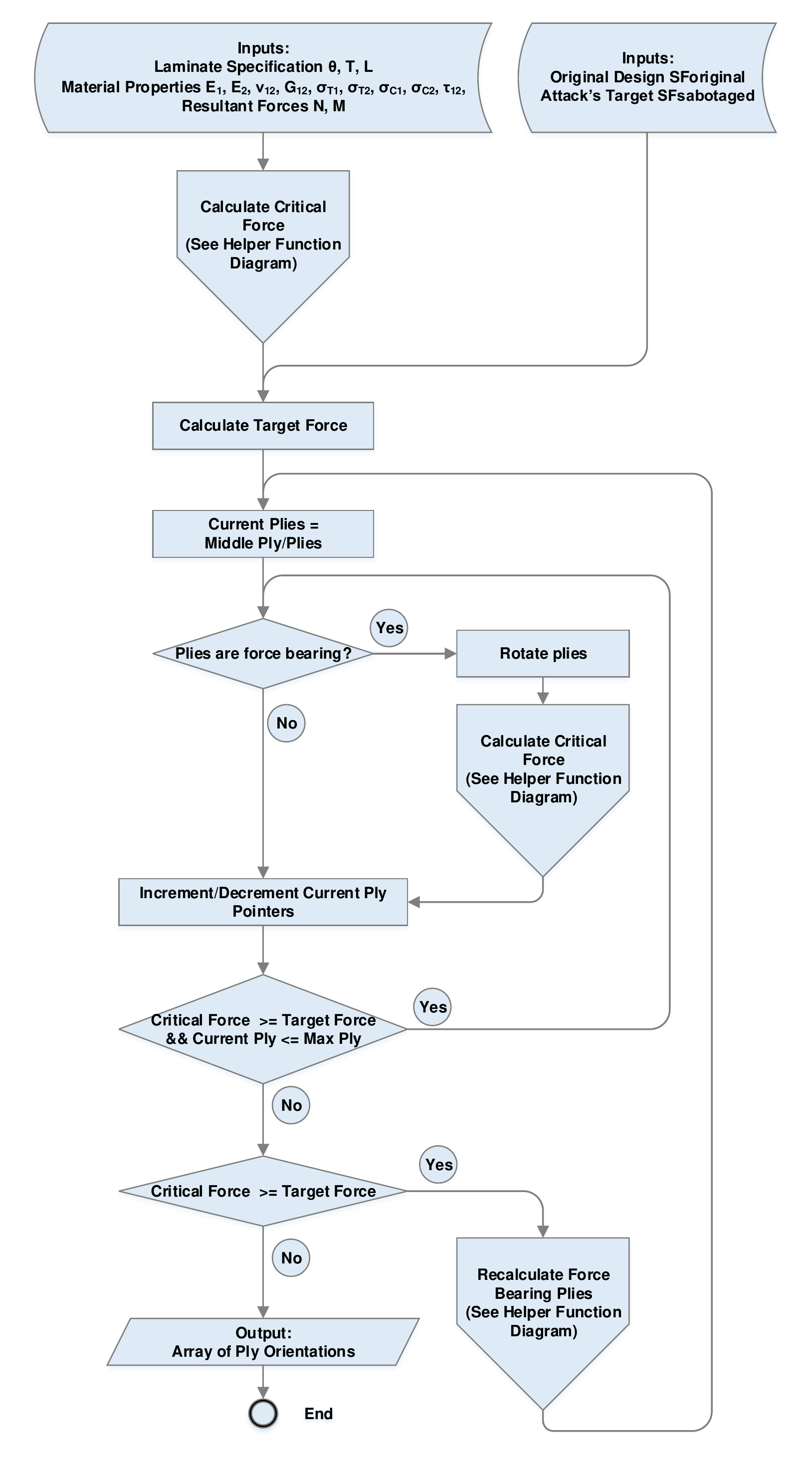}
	\caption{Computation of Type 1 Optimal Sabotage Attack}
	\label{fig:Attack-Type1}
\end{figure}

The algorithm for the first attack type (see Figure~\ref{fig:Attack-Type1}) requires the specification of the original laminate (layer thickness, ply orientation angles, and the number of layers), properties of the materials in the laminate (Young's modulus, Poisson's ration, etc.), and the resultant forces. Further, the algorithm takes into account two safety factors (SF). 
The first SF specifies the ratio by which the original composite design exceeds the expected operational conditions. The target SF describes the desired degradation of composite's mechanical properties as a ratio of the expected operational conditions. 

In the algorithm, based on the specifications of the original design, we calculate the \emph{critical force} that this design can withstand. This force together with both strength ratios is then used to calculate the target critical force, the force under which the sabotaged composite should fail.

We then enter a loop over all of the plies. For each iteration, the algorithm uses two pointers to ``Current Plies.'' These are plies that undergo modifications of their orientation. We initialize these pointers to one or two middle plies (depending on the total amount of plies used in the laminate). Then, we check whether these plies are critical force bearing or not. If they are, we rotate the plies. If the specified ply angle is negative, rotation is going to be in the negative direction; otherwise, we rotate the ply in the positive direction. After rotation, the critical force is recalculated, using the helper function described in Section~\ref{sec:optimal-attacks:helper-function} (see Figure~\ref{fig:OptimalAttackCalculation_HelperFunction}) and the two pointers to current plies are either incremented or decremented, so that they traverse outwards. If the considered plies are not critical force bearing, only the pointers are updated.

If, after the rotation of the current plies, the critical force still exceeds the target and the end of the laminate is not yet reached, we repeat the rotation process for the new current plies. If the target force is reached, the algorithm returns the updated orientation of all plies.


\subsection{Optimal Attack, Type Two: Minimizing Amount of Tampered Plies}

\begin{figure}[tbp] 
	\centering
		\includegraphics[width=0.5\textwidth]{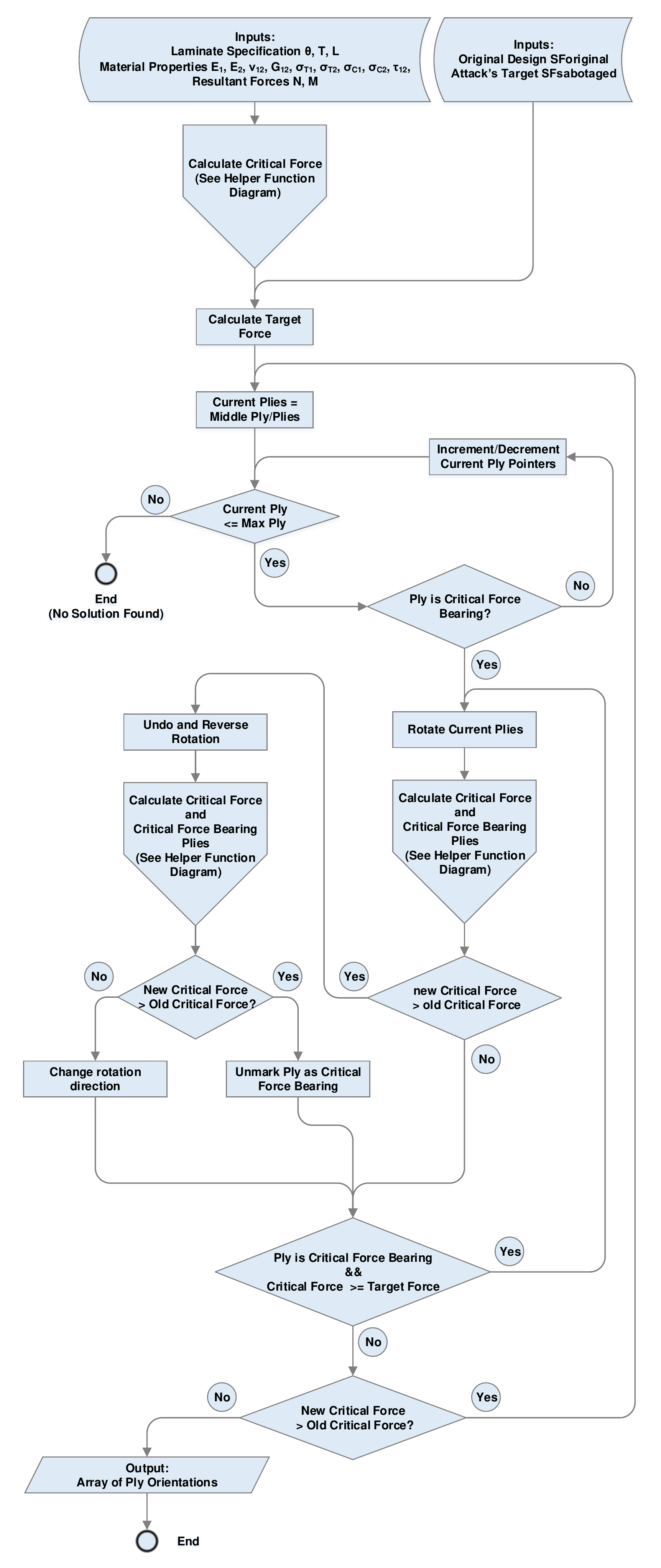}
	\caption{Computation of Type 2 Optimal Sabotage Attack}
	\label{fig:Attack-Type2}
\end{figure}

The algorithm for the second attack type (see Figure~\ref{fig:Attack-Type2}) takes the same parameters as the first. It also performs the same initial calculations to determine the critical and target forces. The current ply initialization is also identical.

Starting with the current plies, the algorithm first tries to find the next critical force bearing plies, traversing the sequence of plies from the center outwards. If the algorithm can't find a critical force bearing ply, this means that no solution could be found. 

Whenever a critical force bearing ply is found, the algorithm begins rotating it. If a 1 degree positive rotation results in a higher critical force, a 1 degree negative rotation is tried. Whichever rotation direction resulted in a lower critical force is repeated until it begins increasing again, indicating a local minimum. If the critical force is still above the target at this point, the next critical force bearing ply is located and undergoes the same rotation process. On reaching the target, the algorithm outputs the updated ply orientations.

\section{Simulation Results and Analysis}
\label{sec:simulation-results}

To analyze a realistic scenario, we use the composite design of a wing's spar proposed in~\cite{kong2007structural} for a small scale 20 seater WIG (Wing-in-Ground Effect) vehicle. The composite consists of 34 layers of Carbon Fiber Reinforced Polymer (CFRP) composite material. 
In all our calculations we assumed a \emph{graphite fiber/epoxy} composite (a.k.a. \emph{carbon fiber/epoxy}). The material properties for graphite fiber/epoxy composite are summarized in Table~\ref{tab:graphite-epoxy-properties}~\cite{kaw2005mechanics}.
For each ply in the composite material, an angle is specified. While the rotation alters the ply's mechanical properties in various directions, all plies combined provide the mechanical properties required for the spar. 

\begin{table}
	\centering
		\begin{tabular}{| l | l |}
			\hline
			\textsc{\textbf{Property}}     & \textsc{\textbf{Value}} \\
			\hline
			
			$E_1$                   & 181  $GPa$ \\
			$E_2$                   & 10.3 $GPa$ \\
			$G_{12}$                & 7.17 $GPa$ \\
			$\nu_{12}$              & 0.28       \\
			${(\sigma^T_1)_{ult}}$  & 1500 $MPa$ \\
			${(\sigma^C_1)_{ult}}$  & 1500 $MPa$ \\
			${(\sigma^T_2)_{ult}}$  & 40   $MPa$ \\
			${(\sigma^C_2)_{ult}}$  & 246  $MPa$ \\
			${(T_{12})_{ult}}$      & 68   $MPa$ \\
			\hline
		\end{tabular}
	\caption{Material Properties for Graphite Fiber/Epoxy Composite (Source:~\cite{kaw2005mechanics})}
	\label{tab:graphite-epoxy-properties}
\end{table}

In our experimental evaluation, we have assumed that the mechanical properties were designed with an SF of 1.5. This is a common approach in the aerospace industry, in order to accommodate for rare extreme conditions during operation as well as for uncertainties in the manufacturing process, including possible impurities of the source materials or deviation in the manufacturing process like curing parameters. 

\begin{table}[tbp] 
	\centering
		\includegraphics[width=0.45\textwidth]{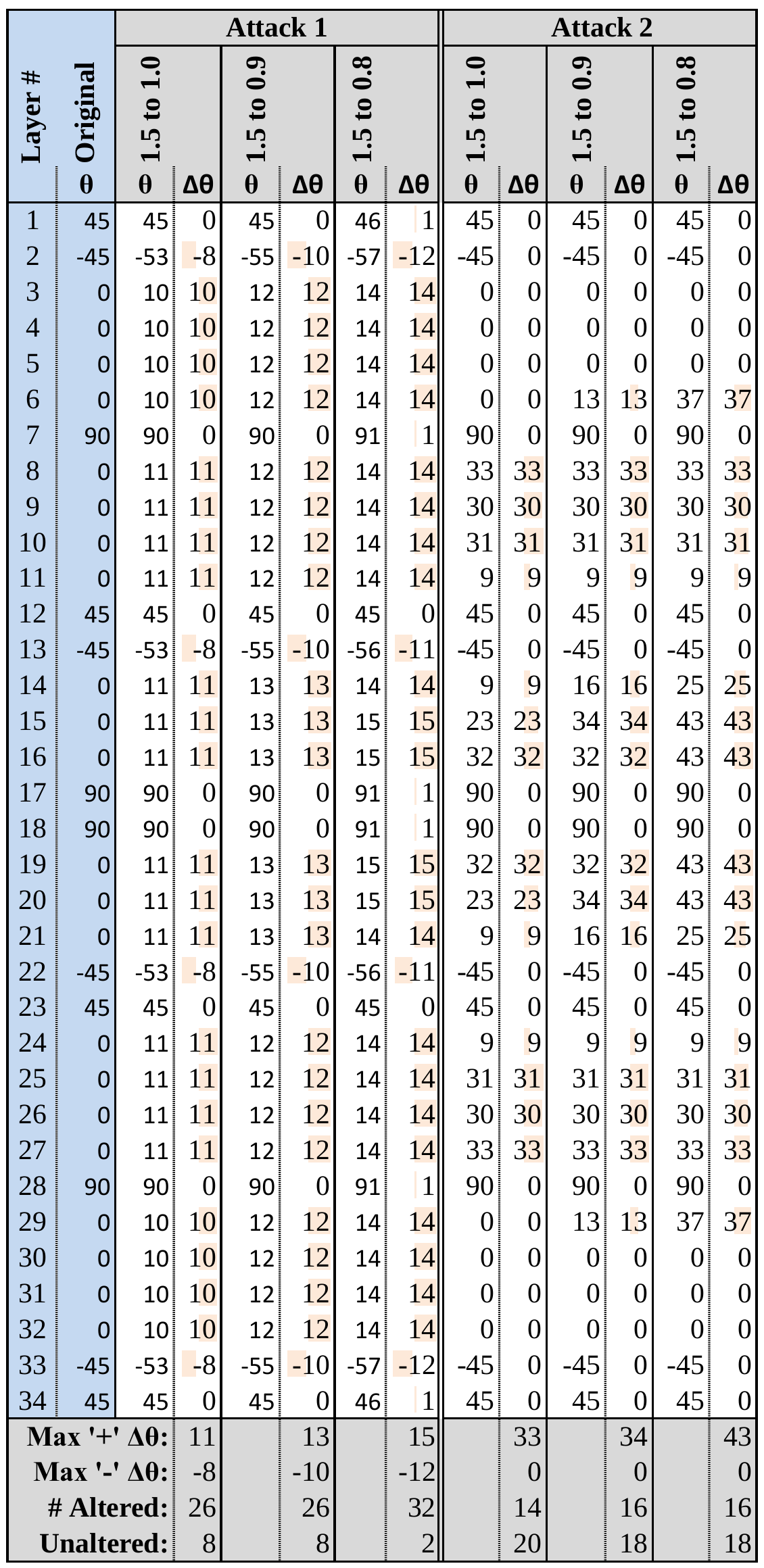}
	\caption{Found Optimal Attacks}
	\label{fig:Table_Attack1-2}
\end{table}

In our simulation, we have used the specification of the composite design and the assumed SF to calculate the forces that the wing's spar should bear during normal operational conditions. Forces and moments should be proportional to the assumed critical forces and moments that will be applied to the wing structure. For the considered spar we use the vector of resultant forces $N=(1, 0, 0) \ N/m$ and resultant moments $M=(0, 0, 0) \ N/m$.

In order to evaluate several attack scenarios, we have simulated attacks that will reduce the SF to three different values: 1.0, 0.9, and 0.8. For both attack types and the three targeted SF, we calculated the necessary changes of the ply orientation. Table~\ref{fig:Table_Attack1-2} summarizes the results of our simulation. It also provides a quick reference for the angle deviation between the altered and the original designs. At the bottom of the table we summarize the maximum angle deviation in both the positive and the negative directions. For each simulated attack, we also provide a number of plies whose orientation was altered or remained unaltered. 

As defined in Section~\ref{sec:optimal-attacks}, a Type 1 attack attempts to achieve the specified mechanical properties degradation while minimizing the angle deviation between the tampered and the original ply orientation. For all considered degradation factors, the deviations are in range between $-12^\circ$ and $15^\circ$. 
Because of the optimization criteria, most plies' orientation must be altered. 
These angles of deviation can be eventually detected during a manual layer inspection; the number of plies affected increase the probability of detection even if not all plies are inspected.

A Type 2 attack is optimized to minimize the number of layers that are altered. For the considered scenario and safety factor degradation, up to 16 layers must be altered in the extreme case. However, this optimization will require a large change in the orientation angle, up to 43 degrees. 
These angles of deviation can be easily detected during the manual inspection. However, if only some layers are inspected, the attack may remain undetected.

\begin{figure*}
    \centering
    \begin{subfigure}[b]{0.5\textwidth}
        \includegraphics[width=\textwidth]{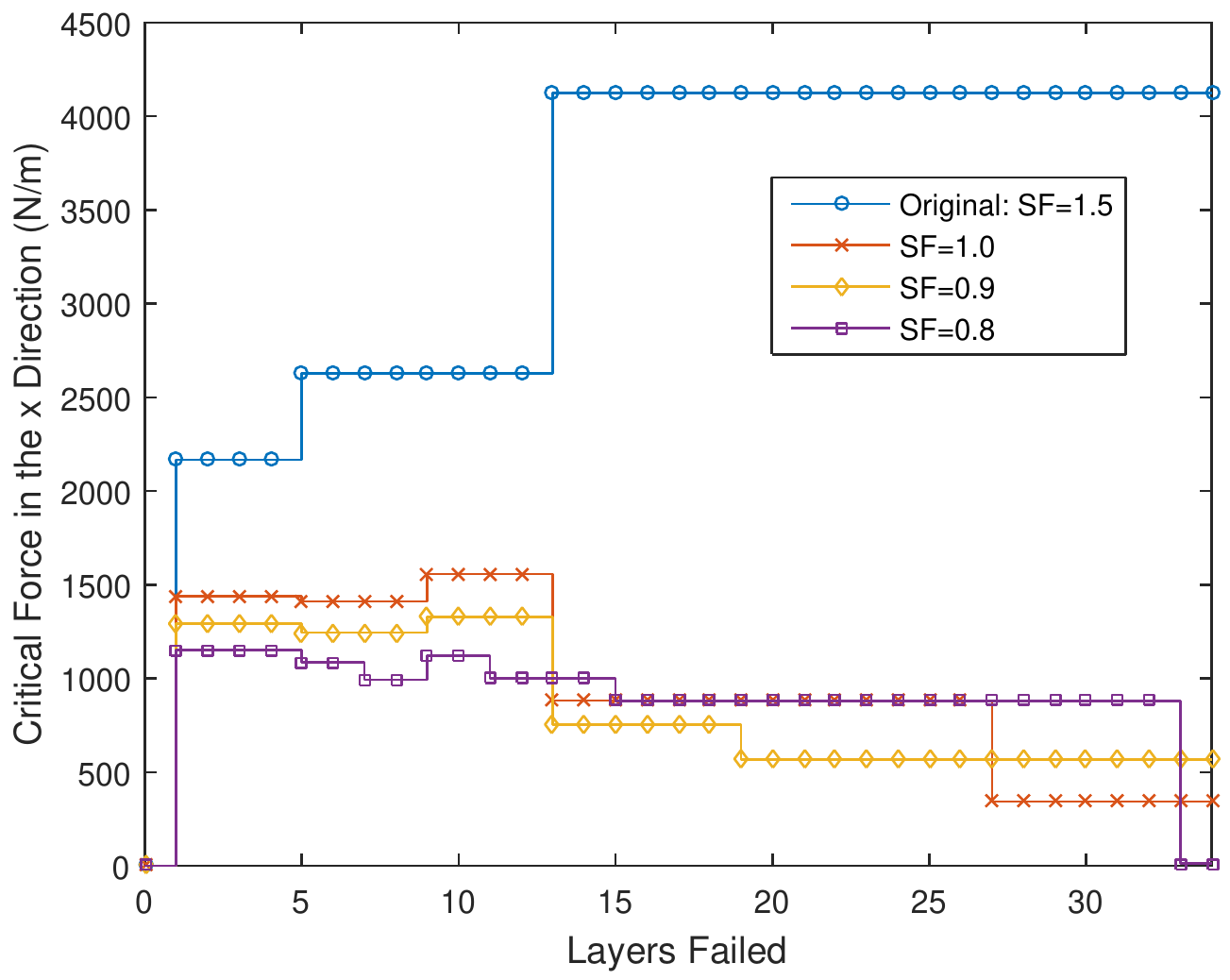}
        \caption{Type 1 Attack}
        \label{fig:SimulationResults_CriticalForces_Attack1}
    \end{subfigure}
		~
    \begin{subfigure}[b]{0.5\textwidth}
        \includegraphics[width=\textwidth]{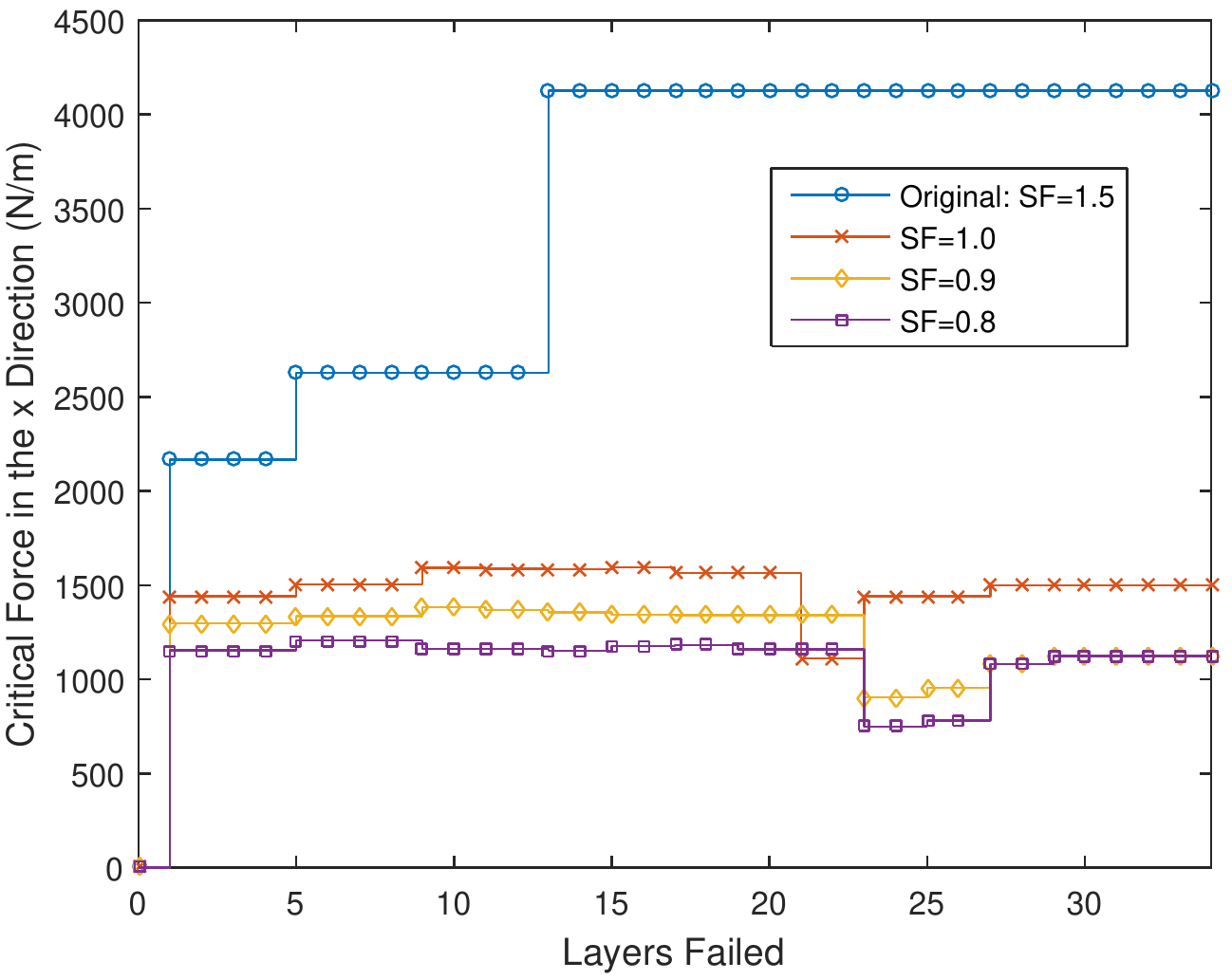}
        \caption{Type 2 Attack}
        \label{fig:SimulationResults_CriticalForces_Attack2}
    \end{subfigure}

    \caption{Simulation Results for Ply Failure}
		\label{fig:SimulationResults_CriticalForces_Attack12}
\end{figure*}

Figure~\ref{fig:SimulationResults_CriticalForces_Attack12}
depict how the attacks alter the critical forces at which plies fail. The original design was optimized to have first ply failure at the applied force of 2166 $N/m$ in the $x$ direction. At this force, only four plies fail. The next failure would occur at a force of 2628 $N/m$ in the $x$ direction and lead to the failure of an additional eight layers. Both of these failures occur well before the entire laminate will fail at a force of 4125 $N/m$ in the $x$ direction. This behavior is characteristic of composite part design and is known as \emph{progressive failure}. The considered design is obviously optimized for safety, even in the event of a first ply failure that can be noticed by pilots or maintenance crews on the ground.

The simulated attacks have fairly different ply failure behavior. Both attacks significantly reduce the forces under which the first ply failure occurs. For all considered scenarios the force is under 1500 $N/m$ in the $x$ direction. Significantly more critical is the follow-up behavior of the sabotaged composite. 
The force increment for the follow-up failure is negligibly small, and it leads to a cascading failure of the remaining plies. Such behavior of a bad composite design is known as \emph{catastrophic failure}. This can obviously lead to catastrophic consequences, such as the crash of an airplane due to the broken spar. 

\section{Identification of Sabotaged Parts}
\label{sec:detectability}

A Ground Vibration Test (GVT) is a non-destructive test commonly performed in aircraft manufacturing. During GVT, data is obtained that can be used to model the flexible dynamics of the tested aircraft. An input force is applied to the aircraft via an electrodynamic shaker. The resulting excitation is measured by accelerometers at multiple points across the aircraft. These excitation points are chosen far away from spar to include both bending and torsion. The measured force and acceleration data obtained is further post processed to determine the modal frequencies and mode shapes using Quadrature Response Method and Curve-Fitting Frequency Domain Decomposition (CFDD)~\cite{gupta2016ground}.

The performed sabotage attacks will change the laminate properties, including its stiffness (see Table~\ref{tab:Table_LaminateStiffness}). The effective orthotropic modulus $E$ can be calculated using the following equation:

\begin{equation} \label{eq:18}
	\frac{1}{E} = \left(\frac{1}{2E_{xx}E_{yy}}\right)^\frac{1}{2} 
	                   \left[  \left(\frac{E_{yy}}{E_{xx}}\right)^\frac{1}{2} - \nu_{yx} + \frac{E_{yy}}{2G_{xy}}
										\right]^\frac{1}{2}
\end{equation}

For isotropic material, the modulus of elasticity has an impact on the fundamental resonance frequency. For a rectangular bar, the relationship is given by the following equation~\cite{tognana2010measurement}:

\begin{equation} \label{eq:19}
	E = 0.9465 
			\left(\frac{mf^2}{b}\right)
			\left(\frac{L^3}{t^3}\right)
			T
\end{equation}

Where $m$ is the mass, $b$ is the width, $L$ is the length, $t$ is the thickness of the sample, and $T$ is the correction factor depend on Poisson's ratio and dimensions of the sample. 
Since the sabotage attacks considered in this paper only alter the laminas' orientation, keeping all remaining parameters constant, the fundamental resonance frequency is directly proportional to the square root of the modulus of elasticity: 

\begin{equation} \label{eq:20}
	f \propto \sqrt{E} 
\end{equation}

Based on this relation, we can calculate the ratio of fundamental resonance frequency change between the original and sabotaged designs ($f/f'$). 
As frequency is one of parameters to measure and analyzed during GVT, its significant changes can be used to detect sabotaged parts. 
Table~\ref{tab:Table_LaminateStiffness} summarizes the impact of the sabotage attacks considered in this paper on this ratio. 

For the first category of sabotage attack and reduction of a safety factor from 1.5 to 1.0 and 0.9, the frequency change is around 10\%. 
For the second category of sabotage attack, the frequency change is significantly lower, and can be as little as 4\%. 
However, taking into account measurement errors and deviations during benign manufacturing process, it is uncertain whether such slight changes of frequency can be (i) detected, and (ii) attributed to a sabotage attack.
Nevertheless, we see changes of fundamental resonance frequency as an important characteristic that could be used to identify sabotaged parts.

\begin{table}
	\centering
		\includegraphics[width=0.5\textwidth]{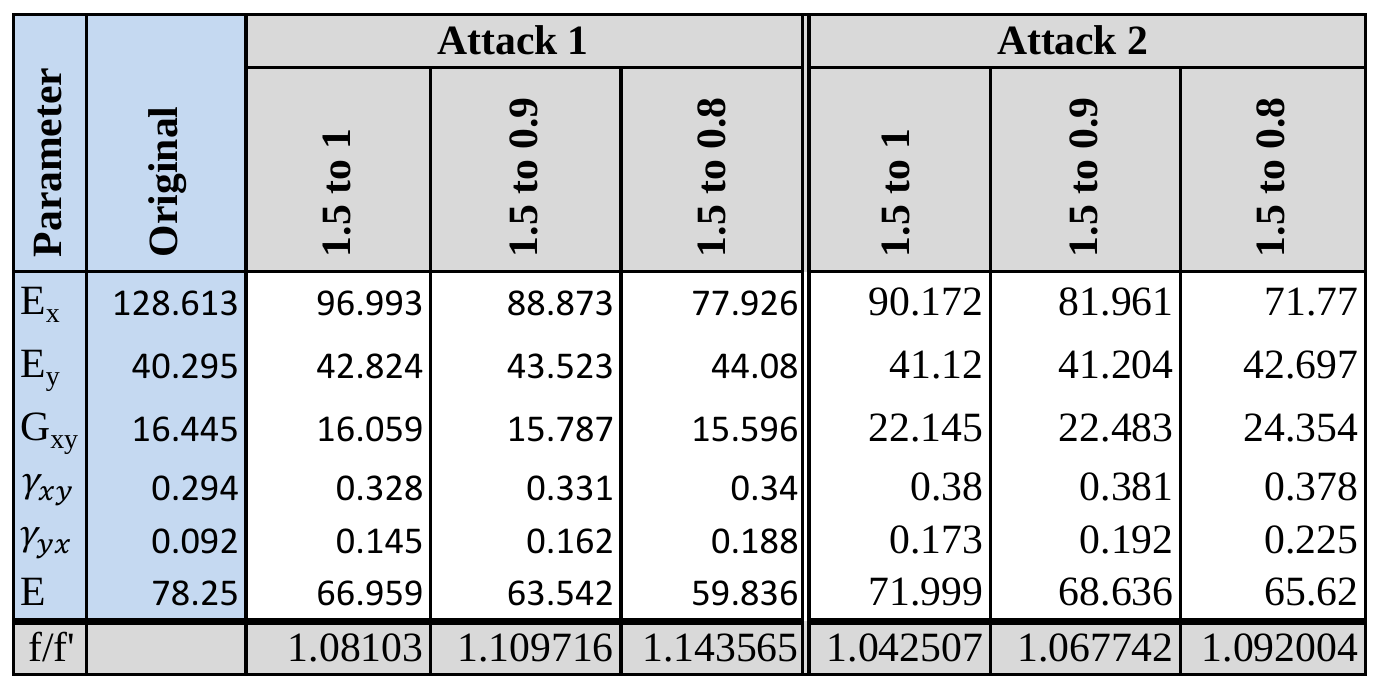}
	\caption{Impact of Sabotage Attacks on Laminate Stiffness 
	and on Fundamental Resonance Frequency}
	\label{tab:Table_LaminateStiffness}
\end{table}

\section{Related Work}
\label{sec:SOTA}

To the best of our knowledge, there are no peer-reviewed publications focusing directly on security of composite materials. There is, however, a growing body of literature in a related field, the security of \emph{Additive Manufacturing} (AM). The identified threat categories for AM include Intellectual Property (IP) violation, manufacturing of legally prohibited objects, and sabotage attacks on manufactured functional parts or manufacturing equipment~\cite{yampolskiy2017evaluation}. For this paper, only sabotage attacks and related literature are of relevance.
It should be noted that, by the end of 2017, no publications have addressed security of composite materials~\cite{yampolskiy2018security}.

Several publications have analyzed 3D printers and the 3D printing processes for vulnerabilities.
Turner et al., 2015~\cite{turner2015bad} found that networking and communication systems lack integrity checks when receiving the design files. 
Moore et al., 2016~\cite{moore2016vulnerability} identified numerous vulnerabilities in software, firmware, and communication protocol commonly used in desktop 3D printers; these can be potentially be exploited.
Do et al., 2016~\cite{do2016data} have shown that communication protocols employed by desktop 3D printers can be exploited, enabling the retrieval of current and previously printed 3D models, halting an active printing job, or submitting a new one.
Belikovetsky et al., 2017~\cite{belikovetsky2017dr0wned} used a phishing attack to install a backdoor that enabled arbitrary, targeted manipulations of design files by a remote adversary.
Sturm et al., 2014~\cite{sturm2014cyber} used malware pre-installed on a computer to automate the manipulation of STL files. 
Moore et al., 2016~\cite{moore2017implications} used malicious firmware to modify and substitute a printed 3D model.

A growing body of publications discusses how a manufactured part's quality can be compromised. 
The majority focus on Fused Deposition Modeling (FDM), commonly used in desktop 3D printers.
Sturm et al., 2014~\cite{sturm2014cyber} demonstrated that a part's tensile strength can be degraded by introducing defects such as voids (internal cavities).
Zeltmann et al., 2016~\cite{zeltmann2016manufacturing} showed that similar results can be achieved by printing part of the structure with the contaminated material.
Belikovetsky et al, 2017~\cite{belikovetsky2017dr0wned} proposed degrading a part's fatigue life; the authors argue that the defect's size, geometry, and location are factors in the degradation. 
Yampolskiy et al, 2015~\cite{yampolskiy2015security} argued that the anisotropy intrinsic to 3D printed parts can be misused to degrade a part's quality, if an object is printed in the wrong orientation.
Zeltmann et al., 2016~\cite{zeltmann2016manufacturing} have experimentally shown the impact of this attack on a part's tensile strength, using 90 and 45 degree rotations of the printed model.
Chhetri et al., 2016~\cite{chhetri2016kcad} introduced a skew along one of the build axes as an attack.
Moore et al., 2016~\cite{moore2017implications} modified the amount of extruded source material to compromise the printed object's  geometry.
Yampolskiy et al.~\cite{yampolskiy2017evaluation}, based on the prior joint work with Pope et al., 2016~\cite{pope2016hazard}, identified that indirect manipulations like the modification of network command timing and energy supply interruptions can be potential means of sabotaging a part. 
Yampolskiy et al, 2015~\cite{yampolskiy2015security} discussed various metal AM process parameters whose manipulation can sabotage a part's quality; for the powder bed fusion (PBF) process, the identified parameters include heat source energy, scanning strategy, layer thickness, source material properties like powder size and form, etc.
Yampolskiy et al., 2016~\cite{yampolskiy2016using} argued that in the case of metal AM, manipulations of manufacturing parameters can not only sabotage a part's quality, but also damage the AM machine, or lead to the contamination of its environment.

Several publications present methods for detecting sabotage attacks.
Chhetri et al., 2016~\cite{chhetri2016kcad} used the acoustic side-channel inherent to the FDM process to detect tampering with a 3D printed object; the authors report that the detection rate of object modifications is 77.45\%.
Belikovetsky et al., 2018~\cite{belikovetsky2018digital} also rely on acoustic emanations of a FDM 3D printer. Authors define five categories of \emph{atomic modifications} at a single G-code command level, and identify the detectability thresholds for each category.
The defined categories are (a) insertion of a command, (b) deletion of a command, (c) modification of movement length on the axis, (d) modification of extruder speed, and (e) re-ordering of G-code commands.
Strum et al., 2017~\cite{sturm2017insitu} proposed an impedance-based monitoring method. The authors physically coupled a piezoceramic (PZT) sensor to the part being fabricated and measure the electrical impedance of the PZT. These impedance measurements can be directly linked to the mechanical impedance of the part, assisting in detecting in-situ defects of part mass and stiffness. 
Two further papers built upon the cross-domain attack notion introduced in Yampolskiy et al., 2013~\cite{yampolskiy2013taxonomy}, and propose a notion of cross-domain attack detection.
Chhetri et al., 2017~\cite{chhetri2017cross} demonstrated the flow of information between the cyber and physical domains and how this information can be used for performing cross-domain security analysis. By estimating this relationship, the model can be used for the detection of new cross-domain attack models and attack detection techniques. 
Wu et al., 2017~\cite{wu2017detecting} leverage machine learning methods to detect cyber-attacks in the manufacturing process. The authors have used vision and acoustics as the data sources for machine learning algorithms and were able to detect anomalies with high accuracy (96.1\% and 91.1\% respectively).

\section{Conclusion}
\label{sec:conclusion}

Composite materials often offer better mechanical properties at lesser weight, as compared to conventional materials. They are popular in the aerospace, high-end automotive, and defense industries. Due to the excellent mechanical properties and ability to optimize these for expected operational conditions, composite materials are frequently used as functional parts in safety-critical systems, including parts of an airplane's wings or fuselage. In order to ensure the quality of composite materials under persistent security threats, it is paramount to understand the properties of potential sabotage attacks.

In this paper, we focused on sabotage attacks that degrade mechanical properties by changing the orientation of individual plies of composite material. We left all other manufacturing parameters unaltered. For this category of sabotage attacks, we have defined two categories of optimal (or minimally invasive) sabotage attacks that degrade mechanical properties to the desired level: (i) minimal amount of deviation from the original ply orientation, and (ii) minimal number of plies whose orientation is changed. 

For the identification of both sabotage attack categories we have developed a MATLAB simulation program that takes as input a sequence of plies, including their material and orientation, and desired degradation; the output of this program, for each attack category, is the changed orientation of plies that would satisfy a specified degradation of mechanical properties. 

We have applied the developed program to the wing spar of a small-scale 20 seater airplane. Assuming a safety factor ($SF$) as 1.5 and desired degraded properties at 1.0, 0.9, and 0.8 of expected operational conditions, we calculated that the necessary deviations can be astonishingly small. For the first category of attacks, it is sufficient to change orientation of all plies by as little as 11 degrees, 13 degrees, or 15 degrees for $SF=1.0$, $SF=0.9$, and $SF=0.8$ scenarios. 
For the second category of attack, it is sufficient to modify the orientation of 16 plies, with maximal deviation of 33 degrees, 34 degrees, and 43 degrees for $SF=1.0$, $SF=0.9$, and $SF=0.8$ scenarios.

While from the cyber perspective it sounds like a significant difference, the ability to detect such attacks is not guaranteed. First of all, many manufacturers still employ manual quality control. If every layer is controlled, especially for the first category of attack, deviations can be small enough to evade detection by human quality control team. If CNC machines like Automated Fiber Placement (AFP) and Automated Tape Layup (ATL) are used, a cyber-attack that changes the design file will both degrade the mechanical properties and compromise the quality control measures. 

Especially critical is the catastrophic failure characteristic of the sabotaged composite. 
After the first ply failure, the force increment for the follow-up failure is negligibly small, and it leads to almost immediate destruction of all remaining plies. This undermines a critical characteristic of composite material design, progressive failure. Such rapid failure could easily lead to disasters, such as airplanes crashing from a failed wing spar.

\bibliographystyle{ACM-Reference-Format}



\balance


\appendix
\section{Calculating Mechanical Properties: Hooke's Law for a Two-Dimensional Angle Lamina}
\label{sec:appendix}
\label{sec:background:math}

Figure~\ref{fig:12XY-Axes} shows the local and global axes of an angle lamina. 1-2 represents the local coordinate axes whereas x-y represents the global coordinate axes. 1 lies along the fiber length direction, which represents longitudinal direction. 2 is normal to the fiber length and represents transverse direction for the lamina. Together they represent the principal material directions in the plane of the lamina. Axis 3, which is transverse to both the fiber axis and the plane of the lamina, is not shown in the figure as the lamina is considered two dimensional. The global coordinate system (x-y) is related to the local coordinate system by the angle ($\Theta$).

Laminas have low strength and stiffness in the transverse direction. So, angled laminas are placed in the laminate to overcome these properties. Fiber reinforced composites are inhomogeneous and non-isotropic (orthotropic) in nature; therefore, Hooke's law (${\sigma=E\epsilon}$) cannot be applied directly. 
The stress-strain equations for orthotropic lamina in both the local and global coordinate systems are given by Equations~\ref{eq:1} and~\ref{eq:2}, respectively~\cite{kaw2005mechanics}.

\begin{equation} \label{eq:1}
	\begin{bmatrix}
		\sigma_1 \\
		\sigma_2 \\
		\tau_{12}	
	\end{bmatrix}
	=
	\begin{bmatrix}
			Q_{11} & Q_{12} & 0 \\
			Q_{21} & Q_{22} & 0 \\
			0      & 0      & Q_{66}
	\end{bmatrix}	
	\begin{bmatrix}
		\epsilon_1 \\
		\epsilon_2 \\
		\gamma_{12}	
	\end{bmatrix}	
\end{equation}


\begin{equation} \label{eq:2}
	\begin{bmatrix}
		\sigma_x \\
		\sigma_y \\
		\tau_{xy}	
	\end{bmatrix}
	=
	\begin{bmatrix}
			\overline{Q}_{11} & \overline{Q}_{12} & \overline{Q}_{16} \\
			\overline{Q}_{21} & \overline{Q}_{22} & \overline{Q}_{26} \\
			\overline{Q}_{16} & \overline{Q}_{26} & \overline{Q}_{66}
	\end{bmatrix}	
	\begin{bmatrix}
		\epsilon_x \\
		\epsilon_y \\
		\gamma_{xy}	
	\end{bmatrix}	
\end{equation}

where ${\sigma}$ ${\epsilon}$ 
are the normal stress and strain; 1,2 are the direction; 
${\tau_{12}}$	and ${\gamma_{12}}$	 
are the shear stress and strain in the plane 1-2. 
${[Q]}$ is the reduced stiffness matrix; x, y are the directions of normal stress and strain; 
${\tau_{xy}}$	and ${\gamma_{xy}}$	 
are the shear stress and strain in the x-y plane; 
${[\overline{Q}]}$ 
is the transformed reduced stiffness matrix.

The elements of the reduced stiffness matrix in Equation~\ref{eq:1} are functions of the material constants and calculated as:

\begin{equation} \label{eq:3}
	Q_{11} = \frac{E_1}{1-\nu_{12}\nu_{21}}, 
	Q_{12} = \frac{\nu_{12}E_2}{1-\nu_{12}\nu_{21}}, 
	Q_{22} = \frac{E_2}{1-\nu_{12}\nu_{21}},
	Q_{66} = G_{12}
\end{equation}

Where $E$ is the Young's modulus; 1,2 are direction; ${G_{12}}$ is the shear modulus in 1-2 plane; ${\nu_{12}}$ and ${\nu_{21}}$ are the Poison's ratios in the 1-2 and 2-1 planes. 

The transformed reduced stiffness matrix ${[\overline{Q}]}$ 
used in the Equation~\ref{eq:2} can be determined by:

\begin{equation} \label{eq:4}
	[\overline{Q}] = [T]^{-1}[Q][R][T] [R]^{-1}
\end{equation}

Where ${[T]}$ and ${[R]}$ are transformation and Reuter matrices which are given as:


\begin{equation} \label{eq:5}
  [T] =
	\begin{bmatrix}
			c^2 & s^2 &  2cs \\
			s^2 & c^2 & -2sc \\
			-sc & sc  &  c^2-s^2
	\end{bmatrix}	
	, and \ 
  [R] =
	\begin{bmatrix}
			1 & 0 & 0 \\
			0 & 1 & 0 \\
			0 & 0 & 2
	\end{bmatrix}	
\end{equation}

Where, ${c=cos(\Theta)}$, ${s=sin(\Theta)}$. 
The relationship between the local stress and strain, from the Equation~\ref{eq:1}, and the  global stress and strain is given by Equation~\ref{eq:6}.

\begin{equation} \label{eq:6}
	\begin{bmatrix}
		\sigma_x \\
		\sigma_y \\
		\tau_{xy}	
	\end{bmatrix}
	= [T]^{-1}
	\begin{bmatrix}
		\sigma_1 \\
		\sigma_2 \\
		\tau_{12}	
	\end{bmatrix}
	, and \ 
	\begin{bmatrix}
		\epsilon_1 \\
		\epsilon_2 \\
		\gamma_{12}	
	\end{bmatrix}	
	= [R][T][R]^{-1}
	\begin{bmatrix}
		\epsilon_x \\
		\epsilon_y \\
		\gamma_{xy}	
	\end{bmatrix}	
\end{equation}

From the Equations above we can determine the stress and strain for a single lamina. 
However, composite laminates are composed of various laminas so Equation~\ref{eq:7} is set up for laminate.

\begin{equation} \label{eq:7}
	\begin{bmatrix}
		N \\
		M
	\end{bmatrix}
	= 
	\begin{bmatrix}
		A & B \\
		B & D
	\end{bmatrix}
	\begin{bmatrix}
		\epsilon^0 \\
		k	
	\end{bmatrix}	
\end{equation}

Where, $N$ is the normal resultant forces (per unit width); $M$ is the bending moment's resultant (per unit width); ${\epsilon^0}$ 
is the mid-plane normal strains; $k$ is the mid-plane curvatures. 
Mid-plane strains and mid-plane curvatures have relation with global coordinate system by Equation~\ref{eq:8}:

\begin{equation} \label{eq:8}
	\begin{bmatrix}
		\epsilon_x \\
		\epsilon_y \\
		\gamma_{xy}	
	\end{bmatrix}	
	=
	\begin{bmatrix}
		\epsilon_x^0 \\
		\epsilon_y^0 \\
		\gamma_{xy}^0	
	\end{bmatrix}	
	+ z
	\begin{bmatrix}
		k_x \\
		k_y \\
		k_{xy}	
	\end{bmatrix}	
\end{equation}

Where $z$ is the distance from the mid-plane in the thickness direction. It is considered positive in the downward direction (see Figure~\ref{fig:LaminatePlyIllistration}). 

The ${[A]}$, ${[B]}$, and ${[D]}$ matrices in the Equation~\ref{eq:7} are called extensional stiffness matrix, coupling stiffness matrix, and bending stiffness matrix for the laminate, respectively. 
The Equations~\ref{eq:9} through~\ref{eq:11} can be used to calculate these matrices.

\begin{equation} \label{eq:9}
	A_{ij} = \sum^n_{k=1}[(\overline{Q_{ij}})]_k(h_k - h_{k-1}) 
\end{equation}
\begin{equation} \label{eq:10}
	B_{ij} = \frac{1}{2}\sum^n_{k=1}[(\overline{Q_{ij}})]_k(h_k^2 - h_{k-1}^2) 
\end{equation}
\begin{equation} \label{eq:11}
	D_{ij} = \frac{1}{3}\sum^n_{k=1}[(\overline{Q_{ij}})]_k(h_k^3 - h_{k-1}^3) 
\end{equation}

Where $n$ is the number of lamina; ${[(\overline{Q_{ij}})]_k}$ 
is the $i$-th, $j$-th element of the 
matrix ${[\overline{Q}}]$ for the $k$-th layer; $i,j = 1,2,3$; $h_k$ is the distance from the laminate midplane to the bottom of the $k$-th lamina. 
The sign convention used for ${h_k}$ is positive below the mid-plane and negative above the mid-plane (see Figure~\ref{fig:LaminatePlyIllistration}).

Among the various failure theories applied to composite structures, \emph{Tsai-Wu} failure theory is considered more general and closely correlated with the experimental data~\cite{kim2014proceedings}. 
According to this failure theory, the lamina is considered to be failed if Equation~\ref{eq:12} is violated.

\begin{equation} \label{eq:12}
	H_1\sigma_1 + H_2\sigma_2 + H_6\tau_{12} + 
	H_{11}\sigma_1^2 + H_{11}\sigma_2^2 + H_{66}\tau_{12}^2 + 
	2 H_{12}\sigma_1\sigma_2 < 1
\end{equation}

The components of the Tsai-Wu failure theory are determined using the five strength parameters of a unidirectional lamina which are given by Equations~\ref{eq:13-1} through~\ref{eq:13-3}:

\begin{equation} \label{eq:13-1}
	H_1 = \frac{1}{(\sigma^T_1)_{ult}} - \frac{1}{(\sigma^C_1)_{ult}}; 
	H_2 = \frac{1}{(\sigma^T_2)_{ult}} - \frac{1}{(\sigma^C_2)_{ult}}; 
	H_6 = 0 \newline
\end{equation}
\begin{equation} \label{eq:13-2}
	H_{11} = \frac{1}{(\sigma^T_1)_{ult} * (\sigma^C_1)_{ult}}; 
	H_{22} = \frac{1}{(\sigma^T_2)_{ult} * (\sigma^C_2)_{ult}}; 
	H_{66} = \frac{1}{(\tau_{12})^2_{ult}}
\end{equation}
\begin{equation} \label{eq:13-3}
	H_{12} = -\frac{1}{2} \sqrt{\frac{1}{(\sigma^T_1)_{ult} * (\sigma^C_1)_{ult} * (\sigma^T_2)_{ult} * (\sigma^C_2)_{ult}}}
\end{equation}

Where ${(\sigma^T_{1,2})_{ult}}$ is the ultimate tensile stress in 1 and 2 direction; 
${(\sigma^C_{1,2})_{ult}}$ is the ultimate compressive stress in 1 and 2 direction; 
${(\tau_{12})_{ult}}$ is the ultimate shear stress in 1-2 plane.

\emph{Strength Ratio} ($SR$) is defined as 

\begin{equation} \label{eq:14}
	SR = \frac{Maximum \ Load \ which \ Can \ be \ Applied}{Load \ Applied}
\end{equation}

When ${SR>1}$, the lamina is safe and the applied stress can be increased. If ${SR<=1}$, the lamina is unsafe and the applied stress needs to be reduced.
For example, ${SR}$ value of 1.8 means that loading can be increased up to 80\% without failure. The concept of SR is applied to any failure theory. 
For the better use of this theory each stress component in Equation~\ref{eq:12} was multiplied by ${SR}$ and resulted an Equation~\ref{eq:15}~\cite{ramsaroop2010using}.

\begin{equation} \label{eq:15}
	(H_1\sigma_1 + H_2\sigma_2 + H_6\tau_{12})SR + 
	(H_{11}\sigma_1^2 + H_{11}\sigma_2^2 + H_{66}\tau_{12}^2 + 2 H_{12}\sigma_1\sigma_2)SR^2 < 1
\end{equation}


\end{document}